\documentstyle[12pt]{article}

\def\p{\partial}
\def\and{\quad{\rm and}\quad}
\def\for{\quad{\rm for}\quad}
\def\with{\quad{\rm with}\quad}
\def\where{\quad{\rm where}\quad}
\def\bra{[\![}
\def\ket{]\!]}
\def\qexp{{\rm Exp}_q}
\def\theorem#1{\medskip\noindent{\bf Theorem\ } #1 :\ }
\def\proposition#1{\medskip\noindent{\bf Proposition\ } #1 :\ }
\def\half{{1\over2}}

\begin{document}
\rightline{preprint-TU-466/FIM-Sept-94}

\begin{center}
{\LARGE\bf The Quantum Group as a Symmetry}\\[1em]

{\bf - The Schr\"odinger equation of  the $N$-dimensional
 $q$-deformed Harmonic Oscillator - }\\[2em]
 Ursula Carow-Watamura${}^{*}$ and
Satoshi Watamura${}^\dagger$\footnote{Research Fellow of the Canon
Foundation in Europe.} \\[1em]

${}^*$ Fakult\"at f\"ur Physik \\
Universit\"at Freiburg\\
D-79104 Freiburg i. Br., Germany\\[1em]

${}^\dagger$ Forschungsinstitut f\"ur Mathematik\\
ETH-Zentrum\\
CH-8092 Z\"urich, Switzerland\\

and \\

Department of Physics\footnote{Parmanent address. \hfill\\
e-mail:
ursula@kaws.coge.tohoku.ac.jp;
watamura@kaws.coge.tohoku.ac.jp}\\
Tohoku University\\
Aobaku, Sendai 980-77, Japan

\end{center}

\begin{abstract}
\medskip
With the aim to construct a dynamical model with quantum group symmetry,
the $q$-deformed Schr\"odinger equation of the harmonic
oscillator on the $N$-dimensional quantum Euclidian space is investigated.
After reviewing the differential calculus on the $q$-Euclidian space,
the $q$-analog of the creation-annihilation operator is constructed.
It is shown that it produces systematically all eigenfunctions of the
Schr\"odinger equation and eigenvalues.
We also present an alternative way to solve the Schr\"odinger equation
which is based on the $q$-analysis.  We represent the Schr\"odinger
equation  by the $q$-difference equation and solve it by using
$q$-polynomials and $q$-exponential functions. The problem of the
involution corresponding to the reality condition is discussed.

\bigskip
\end{abstract}

\eject

\section{Introduction}

\bigskip

It is an interesting question whether we can generalize the known theories to a
more general
framework which may be useful for finding a unified description of
the quantum theory and the gravity.
It is already a longstanding idea that the spacetime structure has to be
modified to achieve this aim, however the question is how this
modification has to be.
The q-deformation can be understood as a new type of deformation,
which is in principle independent of the deformation corresponding to
quantum mechanics.
Thus the possibility of defining the quantum mechanics on a
quantum space is
considered as a step in this direction.
The quantum space we have here in mind is defined as a $q$-deformed
representation space
of the corresponding quantum group, and it is a non-commutative algebra
\cite{[Manin],[FRT]}.
In this sense, we are considering the quantum mechanics on
non-commutative space.

The discovery of the new possibilities related with the q-deformation
strongly motivated the physicists to investigate whether the
quantum symmetry has some non-trivial impact onto the microscopic spacetime
structure \cite{[WorPusz],[WZ],[Wor],[CSSW],[CSW],[Pod]}.
The generalization to the non-commutative geometry is also considered,
using the framework of A. Connes (as for the original work see
\cite{[Connes]}).
It is still not clear in which way the approach via the quantum groups relates
to his framework.
However, with
the quantum group symmetry we get rather easily into the
non-commutative algebra and can construct for example differential calculi,
integrations etc., by using the analogy to the known theory as a guideline.

We take the point of view that the quantum group and
the quantum space are the $q$-deformation of the usual group and space.
Then requiring the ``$q$-correspondence principle'', i.e. in the limit
$q\rightarrow 1$ we recover the known undeformed system,
we can construct the non-commutative analogue of the known objects
such as the quantum Lorentz group \cite{[Wor],[CSSW]},
the quantum Minkowski space
\cite{[CSSW],[sasaki],[Majid],[Kulish1],[Kulish2]},
 the quantum Poincar\'e group
\cite{[Ruegg],[Wess],[ScWeWe],[Majid],[Dobrev]} and many other properties.
As for the kinematical aspects the investigations proceeded rather
quickly.

As a first step to investigate the dynamics, of course the construction
of an appropriate differential calculus is needed.
In fact differential calculi on $q$-spaces have been constructed in a rather
early stage of the investigations in quantum groups \cite{[WorPusz],[WZ]}.
However the investigations with respect to the dynamical features still
contain a lot of untouched problems and only a few aspects are
investigated so far (For an incomplete list of references see refs.[18-33]
).

With the above described motivation the authors have constructed the
differential calculus on the N-dimensional $q$-Euclidian space, i.e. the
differential calculus covariant under the action of the quantum
group $Fun_q(SO(N))$ ref.\cite{[CSW]}.
Although it became a little more complicated than the
one in refs.\cite{[WorPusz],[WZ]} which is based on $A$-type quantum groups,
it has the advantage that it contains the metric $C_{ij}$ which makes
it possible to
define the Laplacian.
 Using this differential calculus we have investigated the Schr\"odinger
 equation corresponding to the
$q$-deformed harmonic oscillator and have computed the ground state energy
as well as the first two excited energy levels.
However in that stage a systematic construction to all energy levels has
been missing.

Recently the investigations on this line were put forward by
several authors \cite{[CWHSW],[HW],[Fio1],[Fio2]}.
Among them, Fiore \cite{[Fio1]}
proposed some kind of `raising` and `lowering` operators which map
the wavefunctions of the $r$-th level to the one of the $(r+1)$th level
and succeeded to generate all energy levels. However,
these operators defined in ref.\cite{[Fio1]} have an explicit
dependence on the energy level, i.e., there is one such operator for each
level
and thus we have an infinite number of them. Due to this feature, these
operators cannot be considered as the $q$-analogue of the
creation-annihilation operator.

Stimulated by the work of ref.\cite{[Fio1]}
we proceeded further our investigations
and found
the creation-annihilation operator of the N-dimensional
$q$-deformed harmonic oscillator which is level-independent, as one
should expect \cite{[CW5]}. This defines
all energy eigenvalues and gives
the expression of the eigenfunctions in terms of creation operators acting
on the ground state.

Since the creation-annihilation operator is given in
terms of differential operators, in principle all eigenfunctions can be
computed.
It is however not so easy to find the explicit form of the wave function as
a $q$-polynomial in the coordinates $x^i$ with this method.
Thus, we also describe an alternative method to construct the eigenfunctions
of the Schr\"odinger equation, which is another new result of our
investigations represented in ref.\cite{[CW5]}.
This method corresponds to the
analytic construction of the wave function in the non-deformed case.
Reducing the Schr\"odinger equation to a $q$-difference equation we solve it
directly by using $q$-polynomials. The above described two constructions
give the same eigenvalues and the resulting wave functions have a one-to-one
correspondence.

\section{$q$-Deformed Differential Calculus}

Here we review the $q$-deformed differential calculus on the quantum
Eucledian space, i.e., the space on which the quantum group
$Fun_q\big(SO(N)\big)$ is acting, obtained in refs.\cite{[CSW]}.
As is known, the $\hat R$-matrix of the quantum group $Fun_q(SO(N))$ can
be written by using projection operators as

\begin{equation}
\hat R=q{\cal P}_S-q^{-1}{\cal P}_A +q^{1-N}{\cal P}_1 ,   \label{(I.1)}
\end{equation}
We are considering generic values of $q$, $0<q<1$, therefore the
decomposition of the tensor product to the irreducible representations is
analogous to the classical case. The projectors
${\cal P}_S$, ${\cal P}_A$ and ${\cal P}_1$ give the components of
symmetric traceless, antisymmetric and singlet tensors, respectively
\cite{[FRT]}.

The quantum group algebra is characterized by the commutation relation
\begin{equation}
\hat R MM=MM\hat R ,     \label{(I.1a)}
\end{equation}
where $M=(M^i_j)$ is a $N\times N$ matrix.

 Making use of this projector
decomposition we defined the differential calculus by the algebra
${\bf C}\!<x^i, dx^i, \p^i>$ with relations consistent with the quantum
group action.
The commutation relations of the coordinate functions $x^i$ are given by
requiring their $q$-antisymmetric product to vanish:
\begin{equation}
{\cal P}_A{}^{ij}_{kl}x^k\,x^l=0 .  \label{(I.2)}
\end{equation}
Correspondingly, the wedge product of the one forms on the quantum space
is defined by requiring the $q$-symmetric components to vanish
\begin{equation}
{\cal P}_S(dx\wedge dx)=0\ ,\quad {\cal P}_1(dx\wedge dx)=0 \  .
\label{(I.4)}
\end{equation}
The exterior derivative is defined as a map from the coordinates to the
one forms:
\begin{equation}
d\quad : \quad x^i\quad \rightarrow\quad dx^i
\end{equation}
with the properties of nilpotency:
\begin{equation}
d^2=0,
\end{equation}
the graded Leibniz rule
\begin{equation}
d(ab)=(da)b+(-1)^{ab}a(db)\ ,
\end{equation}
and covariance under the quantum group action.
This leads us to the following further relation:
\begin{equation}
x^idx^j=q\hat R^{ij}_{kl}dx^k x^l  ,  \label{(I.3)}
\end{equation}
To introduce the differential operator $\p^i$ we postulate the
decomposition of the exterior derivative as
\begin{equation}
d=(dx\cdot \p)=C_{ij}dx^i\p^j ,
\end{equation}
where $C_{ij}$ is the metric of the quantum space and normalized as
\begin{equation}
C^{ij}C_{ij}=Q_N\equiv {(1-q^N)\mu\over (1-q^2)}     ,  \label{(I.9)}
\end{equation}
$C^{ij}$ is the inverse metric and $\mu$ is the factor
\begin{equation}
\mu=1+q^{2-N}    .  \label{(I.10)}
\end{equation}
With this decomposition and the consistency with other relations, we get the
remaining relations:

\begin{equation}
\p^ix^j=C^{ij} +q\hat R^{-1 ij}_{\ \ kl}x^k\p^l  ,  \label{(I.5)}
\end{equation}
\begin{equation}
{\cal P}_{A\ kl}^{\ \ ij}\p^k\p^l=0   ,  \label{(I.6)}
\end{equation}
\begin{equation}
\p^idx^j=q^{-1}\hat R^{ij}_{kl}dx^k\p^l .  \label{(I.7)}
\end{equation}
This completes the algebra of the differential calculus on $N$-dimensional
$q$-Euclidian space.

In this algebra, one finds the natural $q$-analogue of the Laplacian
$\Delta$ :

\begin{equation}
\Delta=(\p\cdot\p)=C_{ij}\p^i \p^j     \ .  \label{(I.11)}
\end{equation}
The existence of the Laplacian was one of our motivations
to investigate the differential calculus on the $q$-Euclidian space.

This algebra has another remarkable structure which was found
in ref.\cite{[OZ]}:
First, there is an element $\Lambda$ which counts the conformal dimensions of
the
algebra elements as
\begin{equation}
\Lambda x^i=q^2 x^i\Lambda\quad {\rm and}\quad \Lambda \p^i=q^{-2}\p^i\Lambda
\end{equation}
where the explicit form of $\Lambda$ is given by
\begin{equation}
\Lambda=1+(q^2-1)(x\cdot\p)+{q^{2-N}(q^2-1)^2\over \mu^2}(x\cdot x)\Delta\ .
\end{equation}
Then extending the algebra by including $\Lambda^{-1}$, one finds
an element $\widehat{\p}^i$ with the property:
\begin{equation}
\widehat{\p}^ix^j=C^{ij}+q^{-1}\widehat{R}^{ij}_{kl}x^k\widehat{\p}^l\ .
\label{(I.hat)}
\end{equation}
The explicit form of this element is
\begin{equation}
\widehat{\p}^i={1\over\mu}\Lambda^{-1}[\Delta,x^i]\ ,
\end{equation}
where $[\cdot,\cdot]$ is the usual commutator.  This element satisfies the
following simple relations:
\begin{equation}
\widehat{\p}^i\p^j=q\widehat{R}^{ij}_{kl}\p^k\widehat{\p}^l\ ,
\end{equation}
\begin{equation}
{\cal P}^{ij}_{A\,kl}\widehat{\p}^k\widehat{\p}^l=0\ .
\end{equation}

On the other hand, it is known \cite{[FRT]} that we can define the
$*$-conjugation by extending the $*$-structure of the quantum group to
a coalgebra antihomomorphism
\begin{equation}
*\ :\quad x^i,\ \p^i\quad \rightarrow
\quad (x^i)^*= x^*_i,\
(\p^i)^*=\p^*_i=-q^{-N}\bar\p_i\ .
\end{equation}
where $\bar\p_i$ is introduced in such a way that the $*$-conjugation of
eq.(\ref{(I.5)}) becomes simpler.
Taking the $*$-conjugation of the eq.(\ref{(I.5)}) and
multiplying $\widehat{R}$ on both sides, we obtain
\begin{equation}
\bar\p^{i}x^{*j}=C^{ij}+q^{-1}\widehat{R}^{ij}_{kl}x^{*k}\bar\p^{l}.
\label{(I.5star)}
\end{equation}

Comparing equations (\ref{(I.hat)}) and (\ref{(I.5star)}), it suggests that
one can define an involution for the extended algebra,
\begin{equation}
x^{*i}\equiv(x^j)^*C^{ji}=x^i\ ,
\end{equation}
\begin{equation}
\bar\p^{i}\equiv -q^N(\p^j)^*C^{ji}=\widehat{\p}^i\ .
\end{equation}
In ref.\cite{[OZ]} the authors proposed to take this as an extension of
the reality condition.
Note that under this conjugation the invariant
``length'' $(x\cdot x)=C_{ij}x^ix^j$ and the Laplacian $\Delta=(\p\cdot\p)$
transform as
\begin{equation}
(x\cdot x)^*=(x\cdot x)\quad {\rm and}\quad
(\p\cdot\p)^*=q^{-2N}(\bar\p\cdot\bar\p)=q^{-N-2}\Lambda^{-1}\Delta
\end{equation}

\section{The $q$-Deformed Schr\"odinger Equation}

\subsection{The $q$-deformed harmonic oscillator }

The $q$-deformed Laplacian of the differential calculus on the N-dimensional
$q$-Euclidian space led us to investigate the corresponding Schr\"odinger
equation, the simplest example of which is the harmonic oscillator. The
action of its Hamiltonian onto the wave function $\big|\Psi\big>$ is defined
by
\begin{equation}
H(\omega)\big|\Psi\big>\, =[-q^{N}(\p\cdot\p)+\omega^2(x\cdot
x)]\big|\Psi\big>\, =\, E\big|\Psi\big>\ .
\label{(I.12)}
\end{equation}

We solved this $q$-deformed Schr\"odinger equation for the ground state
 and the first two excited energy
levels by using an appropriate anzatz in ref.\cite{[CSW]}.
The ground state wave function is given by the $q$-exponential function as
\begin{equation}
\big|\Psi_0\big>\,={\rm exp}_{q^2}\bigg[{-\omega (x\cdot x)\over q^N\mu}\bigg]
\ .
\label{(I.13)}
\end{equation}
For the definition of the
$q$-exponential function and some of its
properties see the appendix of ref.{[CW5]}.
The ground state energy is

\begin{equation}
H(\omega)\big|\Psi_0\big>\,=E_0\big|\Psi_0\big> , \quad{\rm where}\quad
E_0={\omega\mu(1-q^N)\over
(1-q^2)}=\omega Q_N   .  \label{(I.14)}
\end{equation}
For the first excited level we have the eigenfunction of the vector
representation $\big|\Psi^i_{1}\big>$:
\begin{equation}
H(\omega)\big|\Psi^i_1\big>\,=E_1\big|\Psi^i_1\big> , \with
E_1={\omega\mu(1-q^{N+2})\over q (1-q^2)}
,  \label{(I.15)}
\end{equation}
where
\begin{equation}
\big|\Psi^i_{1}\big>=x^i {\rm exp}_{q^2}
\bigg[{-\omega (x\cdot x)\over q^{N+1}\mu}\bigg]\ ,
\end{equation}
and for the second excited level the symmetric tensor
$\big|\Psi_{2,S}\big>$ and singlet representation $\big|\Psi_{2,1}\big>$
with the same energy eigenvalue $E_2$:

\begin{eqnarray}
H(\omega)\big|\Psi^{ij}_{2,S}\big>\,&=&E_2\big|\Psi^{ij}_{2,S}\big> ,
\label{(I.16)}\\
H(\omega)\big|\Psi_{2,1}\big>\,&=&E_2\big|\Psi_{2,1}\big> ,  \with
E_2={\omega\mu(1-q^{N+4})\over
q^2 (1-q^2)}\ . \label{(I.17)}
\end{eqnarray}
The corresponding wave functions are given by
\begin{equation}
\big|\Psi_{2,S}\big> = {\cal P}_S(x\otimes x) {\rm exp}_{q^2}
\bigg[{-\omega (x\cdot x)\over
q^{N+2}\mu}\bigg]\ ,
\end{equation}
and
\begin{equation}
 \big|\Psi_{2,1}\big> = \big((x\cdot x)
-{Q_N q^2\over \omega(1+q^2)}\big){\rm exp}_{q^2}
 \bigg[{-\omega (x\cdot x)\over  q^{N+2}\mu}\bigg]\ .
\end{equation}

As for the $q$-numbers we use the conventions:
\begin{equation}
\bra x\ket={q^x-q^{-x}\over q-q^{-1}}   , \label{(I.18)}
\end{equation}
and
\begin{equation}
(x)_{q^2}={1-q^{2x}\over 1-q^2}  . \label{(I.19)}
\end{equation}
Note that the energy levels of the $q$-deformed hamornic oscillator are not
equidistant and also the factors in the $q$-exponentials are different for
different energy levels.

\subsection{The operator formalism}
\medskip

Ordinary quantum mechanics suggests
to look for the creation
and annihilation operator $a^i$ which satisfies in general
a commutation relation with the Hamiltonian of the following form:

\begin{equation}
H(\omega) a^i=q^k a^i[H(\omega)+C(\omega)]\ ,
\label{(O11)}
\end{equation}
where we introduced a possible $q$-factor $q^k$ ($k$ is a real constant).
Such a commutation relation should then generate all eigenfunctions of the
$q$-deformed Schr\"odinger equation.
Then, $a^i$ maps the $p$th state to the $(p+1)$th state and the constant
$C(\omega)$ gives the energy difference between the two states.  However in
a $q$-deformed system there is not such an operator.
The reason is that the energy difference between the neighbouring states is
not equidistant as we see from the eigenvalues
$E_0$, $E_1$ and $E_2$. To account for this feature the author of
ref.\cite{[Fio1]} introduced operators separately for each state which raise
and lower
the energy level, i.e., the $p$-th raising operator $a^\dagger_p$ acts as :
$|p>\rightarrow |p+1>$ (and correspondingly the lowering operator $a_p$) for
the $p$th level.  In this way he obtains all the states together
with an infinite number of raising (and lowering) operators.
 It is a priori not obvious whether one can define at all a
 creation-annihilation operator which produces all states of this
 $q$-deformed system.

The key point to find the creation-annihilation operator of this system is
to allow the quantity $C(\omega)$ to be a function of the Hamiltonian.  One
can easily see that with this generalization the operator $a^i$ still maps
one eigenstate to another eigenstate of different energy level. Using the
analogy with the non-deformed case we look for a creation-annihilation
operator of the form  $(\p^i+x^i\alpha)$ and the above generalization means
that the coefficient $\alpha$ is a function of the Hamiltonian.

In our construction we also have to take into account that the coordinate
function $x^i$ and the derivative $\p^i$ have non-trivial commutation
relations with the Hamiltonian:

\begin{eqnarray}
\p^iH(\omega) &=&H(q\omega)\p^i+\mu\omega^2x^i \ , \label{(2.6.2)}\\
x^iH(\omega)&=&q^{-2}H(q\omega)x^i+q^{N-2}\mu\p^i \ , \label{(2.7.3)}
\end{eqnarray}
With the above described considerations we found the following operators:

\bigskip

\theorem{A}

\noindent{\bf i)}
The creation operator $a^i_-$ and annihilation operator $a^i_+$ are
defined by
\begin{equation}
a^i_{\pm}=q^{-\half}\lambda^{-\half}[q^{N\over2}\p^i+x^i\alpha_\pm(\omega)]\
,
\label{(2.19.4)}
\end{equation}
where
\begin{equation}
\alpha_\pm(\omega)=\half[KH(\omega)\pm\sqrt{K^2[H(\omega)]^2+4\omega^2}]
\with K={(1-q^2)\over q^{N\over2}\mu} \ ,  \label{(2.11.5)}
\end{equation}
and
\begin{equation}
\lambda^{-\half}x^i=q^{-\half}x^i\lambda^{-\half}\and
\lambda^{-\half}\p^i=q^{\half}\p^i\lambda^{-\half}\ .
\label{(2.xx.6)}
\end{equation}

\noindent{\bf ii)}
The commutation relation of the creation and annihilation operators
with the Hamiltonian is
\begin{equation}
H(\omega)a^i_\pm =q^{-1}a^i_\pm [H(\omega)-q^{{N\over2}}\mu
\alpha_\pm(\omega)]\ .  \label{(2.23.7)}
\end{equation}

\bigskip

The all proofs of the theorems and propositions given here are published in
ref.\cite{[CW5]}.
Note that the shift operator $\lambda$ relates to the algebra element
$\Lambda$ mentioned previously as $\lambda^2=\Lambda$.

With these operators defined in Theorem A
we can derive the whole set of states of the corresponding $q$-deformed
Schr\"odinger equation.
We
only need to know the eigenvalue of the ground state.
As we have shown in the ref.\cite{[CSW]}, the wavefunction corresponding to
the ground state in the limit $q\rightarrow1$ is given by
$\big|\Psi_0\big>$ in
eq.(\ref{(I.13)}) and it is a candidate of the ground state for the
$q$-deformed case.
We can prove that the operator $a^i_+$ annihilates
this $\big|\Psi_0\big>$:

\proposition{A}
\begin{equation}
a^i_+\big|\Psi_0\big>=0     \ .         \label{(2.kk.8)}
\end{equation}

The proof makes use of the fact that the action of the operator
$\alpha_{\pm}$ onto the ground state is given by
\begin{equation}
\alpha_\pm(\omega)\big|\Psi_0\big>
=\bigg\{ \matrix{q^{-{N\over2}}\omega\big|\Psi_0\big> &\for &\alpha_+& \cr
                            &&&\cr
		-q^{{N\over2}}\omega\big|\Psi_0\big>&\for&\alpha_- &.\cr} \
\label{(2.31.9)}
\end{equation}
Knowing this we act with the annihilation operator $a^i_+$ onto $|\Psi_0>$
and obtain

\begin{equation}
a^i_+\big|\Psi_0\big>\,
=q^{-\half}\lambda^{-\half}[q^{N\over2}\p^i
-q^{-{N\over2}}\omega x^i]\big|\Psi_0\big>=0 \ .
\label{(2.32.10)}
\end{equation}
On the other hand, eq.(\ref{(2.32.10)}) actually defines the ground state wave
function.

The excited states are obtained by successively applying the creation
operator $a^i_-$
onto the ground state, the energy eigenvalue of which is defined by
eq.(\ref{(I.14)}).
The energy spectrum can be derived by using eq.(\ref{(2.23.7)}) which
gives the recursion formula of the energy levels from $E_p$ to $E_{p+1}$
when both sides are evaluated on the state $|\Psi_p\big>$.
 After some calculation we
get the energy eigenvalue of the $p$-th level as:

\proposition{B}
\begin{equation}
E_p={\omega\mu\over q^{1-N/2}}\bra{N\over 2}+p\ket \ .   \label{(2.YY.11)}
\end{equation}

Another result which is obtained in the course of deriving the
energy eigenvalue is the value of the operator $\alpha$ when acting on the
state $|\Psi_p>$:

\begin{equation}
\alpha_\pm(\omega)\big|\Psi_p\big>\,
=\bigg\{\matrix{q^{-p-{N\over2}}\omega\big|\Psi_p\big>&\for &\alpha_+&\cr
&&&\cr
-q^{p+{N\over2}}\omega\big|\Psi_p\big>&\for &\alpha_-& .\cr}
\label{(2.xs.12)}
\end{equation}

Thus as a consequence, we obtain the equation
\begin{eqnarray}
a^i_\pm\big|\Psi_p\big>\,
&=&q^{-\half}\lambda^{-\half}[q^{N\over2}\p^i
+x^i\alpha_\pm(\omega)]\big|\Psi_p\big>\nonumber\\
&=&q^{-\half}\lambda^{-\half}
[q^{N\over2}\p^i\pm x^iq^{\mp(p+{N\over2})}\omega]
\big|\Psi_p\big> \ . \label{(2.33.13)}
\end{eqnarray}
Therefore, when the creation-annihilation operator defined in
eq.(\ref{(2.19.4)}) is acting onto an eigenstate and we evaluate only the
operator $\alpha_\pm(\omega)$ then the resulting expression becomes
level-dependent and coincides with the raising operator
constructed by Fiore.

One of the important relations to characterize the operators
$\alpha_\pm(\omega)$ and $a^i_\pm$ are the following commutation relations
which, in spite of the complicated expression for the operator
$\alpha_{\pm}(\omega)$, have a remarkable simple form

\proposition{C}

\begin{eqnarray}
\alpha_\pm(\omega)a^i_+&=&q^{\pm1}a^i_+\alpha_\pm(\omega)\
,\label{(2.34.14)}\\
\alpha_\pm(\omega)a^i_-&=&q^{\mp1}a^i_-\alpha_\pm(\omega)\
,\label{(2.35.15)}
\end{eqnarray}
and
\begin{equation}
\lambda^{-\half}\alpha_\pm(\omega)=q\alpha_\pm(\omega/q)\lambda^{-\half}\
.\label{(2.xxx.16)}
\end{equation}

\subsection{Construction of the wave function}

With the above operator relations we can easily derive the general form
of the wave function for arbitrary energy level. For this end first note
that

\theorem{B}

The $q$-antisymmetric product of the creation (annihilation)
operator vanishes
\begin{equation}
{\cal P}_A(a^i_-a^j_-)=0\quad {\rm and}\quad
{\cal P}_A(a^i_+a^j_+)=0\ .
\label{(2.50.17)}
\end{equation}

Theorem B means that the creation operator satisfies the same commutation
relation as the $q$-space coordinate function $x^i$. Thus for example
\begin{equation}
(a_-\cdot a_-) a^i_-=a^i_-(a_-\cdot a_-)\ ,
\label{(2.52.19)}
\end{equation}
where $(a_-\cdot a_-)=C_{ij}(a^i_- a^j_-)$.
Consequently any state constructed by successively applying the creation
operator $a_-^i$ onto the ground state $\Psi_0$ is a $q$-symmetric tensor.
Thus we may call the creation-annihilation operator $a_{\pm}^i$ a
$q$-bosonic operator.
Let us state the above results as a theorem.

\theorem{C}

The states of the $p$th level constructed by $p$ creation operators $a^i_-$
\begin{equation}
\big|\Psi_p^{i_1i_2\cdots i_p}\big>\,\equiv a_-^{i_1}a_-^{i_2}\cdots
a_-^{i_p}\big|\Psi_0\big>\ ,
\label{(C1)}
\end{equation}
have the energy eigenvalue $E_p=\omega\mu
q^{{N\over2}-1}\bra{N\over2}+p\ket$ and are $q$-symmetric tensors, i.e.,
$\forall{l}\in \{1,...,p\!-\!1\}$
\begin{equation}
{\cal P}_A{}^{jj'}_{i_li_{l+1}}\big|\Psi_p^{i_1\cdots i_li_{l+1} \cdots
i_p}\big>\,=0\ .
\label{(C2)}
\end{equation}

Since the energy eigenvalue depends only on the number of creation
operators, the wavefunctions defined in (\ref{(C1)}) have the same
energy eigenvalue $E_p$ for a fixed level $p$.
The second part of Theorem C is a direct consequence of Theorem B.

Thus the number of states of the $p$th level is equal to the number of
states of the non-deformed case, i.e., ${\Big({N+p \ \atop p}\Big)}$.  The
$q$-symmetric tensor can be split into symmetric traceless tensors
corresponding to the irreducible representations of $Fun_q(SO(N))$ :

\begin{equation}
Sym\Big(\underbrace{N\otimes N\otimes\cdots\otimes N}_p\Big)
=S_p\oplus
S_{p-2}\oplus\cdots\oplus \Big\{\matrix{ S_1&{\rm for\  odd}\ p \cr
		& \cr
		S_0&{\rm for\ even}\ p } \Big\} \ .\label{IRRED}
\end{equation}
Correspondingly the wave function in eq.(\ref{(C1)}) is split into
irreducible components.
With this operator method,
Theorem C defines in principle all eigenfunctions.
However it is not straightforward to obtain an expression of the
eigenfunctions as $q$-polynomials in $x^i$.

\subsection{The wave function as a $q$-polynomial in $x$}

In order to obtain an expression of the eigenfunctions in terms of
$q$-polynomials in the coordinate, a simple way is to use the relations of
the $q$-differential calculus with the $q$-analysis \cite{[Exton]}.

The $q$-symmetric traceless $p$th tensor representation $S_p^I$ can be
constructed as follows:
\begin{equation}
S^I_p
\equiv S^{i_1\cdots i_p}_p(x)= S_p{}^{i_1\cdots i_p}_{j_1\cdots
j_p}x^{j_1}\cdots x^{j_p} \ ,  \label{(2.53.20)}
\end{equation}
where
\begin{eqnarray}
{\cal P}_A{}^{kl}_{i_ji_{j+1}}S_p^{i_1\cdots i_ji_{j+1}\cdots
i_p}(x)&=&0 \ , \label{(2.35.21)}\\
C_{i_ji_{j+1}}S_p^{i_1\cdots i_ji_{j+1}\cdots i_p}(x)&=&0 \ ,
 \label{(2.36.22)}
\end{eqnarray}
for $j=1,...,p-1$.

Since the tensor structure is defined by these $q$-symmetric tensors, the
wave functions can be written by the product of a $q$-symmetric tensor and
a function of $x^2$. Thus to solve the Schr\"odinger equation (\ref{(I.12)}),
we take the ansatz:
\begin{equation}
\big|\Psi\big>\,=S^I_pf(x^2)\ .
\label{(P1)}
\end{equation}
The problem is to fix the function $f(x^2)$.

For this end
we compute the action of the Hamiltonian onto the wave function (\ref{(P1)}).
This lead to the following form of the Schr\"odinger equation:

\begin{eqnarray}
(-q^N\Delta+\omega^2 x^2) S^I_p
f(x^2)&=&\mu^2S_p^I[-q^{2N+2p}x^2D^2_{x^2}
-q^N({N\over2}+p)_{q^2}D_{x^2}+{\omega^2\over\mu^2}x^2]f(x^2)\nonumber\\
   &=&E S_p^If(x^2) \ , \label{(e.27)}
\end{eqnarray}
where the definition of $D_{x^2}$ is
\begin{equation}
D_{x^2}f(x^2)={f(q^2 x^2)-f(x^2)\over x^2(q^2-1)} \ .   \label{(d.28)}
\end{equation}
which is the $q$-difference operator.
Therefore eq.(\ref{(e.27)}) is rewritten as a $q$-difference equation
for $f(x^2)$
\begin{equation}
F(D_{x^2})f(x^2)
=[-q^{2N+2p}x^2D^2_{x^2}-q^N({N\over2}+p)_{q^2}D_{x^2}
+{\omega^2\over\mu^2}x^2-{E\over\mu^2}]f(x^2)=0\ .\label{(DE)}
\end{equation}
To solve this equation we take an ansatz with the $q$-exponential function
since we already know from our previous considerations that the wave
function as to be of such a form.
\begin{equation}
f(x^2)=\sum b_s\qexp(q^{-2s}\alpha)\where \alpha=q^{-N-p}\omega \ .
\label{(DE1)}
\end{equation}
A lengthy but straightforward calculation yields
\begin{eqnarray}
F(D_{x^2})f(x^2)
&=&{q^{{N\over2}-1}\omega\over\mu}
\sum\Big[-b_{s+1}q^{{N\over2}+p}[\![2s+2]\!]\nonumber\\
&\quad&\qquad+b_s\Big([\![{N\over2}+p+2s]\!]
-{E\over q^{{N\over2}-1}\mu\omega}\Big)\Big]\qexp(q^{-2s}\alpha)\ .
\label{(DF)}
\end{eqnarray}

Similar to the case of the non-deformed oscillator,
we look for the solution which has a finite number of terms in the expansion
 eq.(\ref{(DE1)}). In such a case the argument of the $q$-exponential
 functions in
eq.(\ref{(DE1)}) can be shifted to the $q$-exponential of the largest $s$.
Then such a function $f(x^2)$
becomes a polynomial of $x^2$ multiplied with the $q$-exponential which
has a smooth finite limit under $q\rightarrow1$.
On the other hand since the $q$-exponentials with different arguments
generate different powers in $x^2$, they are independent and cannot cancel
each other.  Thus solving the equation, we require that for each term in the
series eq.(\ref{(DF)}) the sum of the coefficients of different exponential
functions separately vanishes.

First we see that in the term for $\qexp(\alpha)$, i.e. for $s=-1$ in
eq.(\ref{(DF)}), the coefficient
of $b_0$  is zero. Therefore we can consistently set $b_s=0$ for $s<0$ and
require that the first nonzero term starts with the $b_0$ term.

Now we require that the series has only a finite number of terms.  To
satisfy this, the second term under the sum, i.e., the coefficient of
$\qexp(q^{-2s}\alpha)$ containing the factor $b_s$
must vanish for a certain $s$.  Calling this largest integer $s$ as $r$,
this requirement defines the energy eigenvalue $E$ as

\begin{equation}
E=q^{{N\over 2}-1}\mu\omega[\![{N\over 2}+p+2r]\!] \ .
\label{(2.54.29)}
\end{equation}

With this eigenvalue, we can set the $b_s=0$ for $s>r$ and can solve the
equation (\ref{(DE)}).
{}From the condition that the sum of all terms with the same argument in the
exponential function vanishes we get the recursion formula

\begin{equation}
b_s=
b_{s+1}{-q^{{N\over2}+p}[\![2s+2]\!]\over[\![{N\over2}+p+2r]\!]
-[\![{N\over2}+p+2s]\!]} \ .
\label{(2.56.30)}
\end{equation}
Therefore we obtain

\begin{equation}
b_s= b_r\prod_{s\le t\le r-1} {q^{{N\over2}+p}[\![2t+2]\!]\over
[\![{N\over2}+p+2t]\!]-[\![{N\over2}+p+2r]\!]} \ .
\label{(coeff)}
\end{equation}
The $b_r$ simply shows the freedom of the overall normalization and thus the
wave
functions are now defined in terms of the $q$-exponential functions and the
$q$-polynomials of the coordinate functions $x^i$ with a finite number of
terms and with the $b_s$ given above as

\begin{equation}
\big|\Psi_{p,r}^I\big>\,=S_p^I\sum_{s=0}^{r} b_s\qexp(q^{-2s}\alpha)\where
\alpha=q^{-N-p}\omega \ .\label{(DE2)}
\end{equation}

The energy eigenvalue of $|\Psi_{p,r}>$ is given by $E$
in eq.(\ref{(2.54.29)}) and it coincides with the
eigenvalue defined by using the creation-annihilation operator in the
previous section. We can also confirm that there is a one-to-one
correspondence between the wave function given in eq.(\ref{(C1)}) and the
one in eq.(\ref{(DE2)}):

The wave function derived in eq.(\ref{(DE2)}) shows that for each $p$th rank
tensor we have an infinite tower of eigenfunctions labeled by the integer
$r$ with the eigenvalue $E_{p+2r}=q^{{N\over 2}-1}\mu\omega[\![{N\over
2}+p+2r]\!]$.  This means that for the fixed eigenvalue $E_{p'}$ there is
one  eigenfunction of $p$th rank tensor for each $p$ which satisfies
$p+2r=p'$ with a positive integer $r$. This is the result given in
eq.(\ref{IRRED}).

This completes the $q$-analytic construction of the eigenfunctions which
gives the $q$-polynomial representation of the wave function corresponding
to the irreducible representations of the $Fun_q(SO(N))$.

\subsection{Algebra of the creation-annihilation operators}

It is interesting to ask how far we can make the analogy of the
operator algebra using the creation-annihilation operator.
We discuss
here some properties of the creation-annihilation operator given in
eq.(\ref{(2.19.4)}).

First we have an alternative representation of the creation-annihilation
operator $b^i_\pm$:

\theorem{D}
The creation-annihilation operator can be also represented by

\begin{equation}
b_\pm=\lambda^{-\half}[q^{N\over2}\p^i+\alpha_\pm(q\omega)x^i]\ ,
\label{ThD0}
\end{equation}
where the operator $\alpha(\omega)$ is given by eq.(\ref{(2.11.5)}). These
two representations satisfy the identity
\begin{equation}
{1\over\sqrt{K^2[H(\omega)]^2+4\omega^2}}b_\pm
=q^{3\over2} a_\pm{1\over\sqrt{K^2[H(\omega)]^2+4\omega^2}}\ . \label{ThD}
\end{equation}

In this representation now the operator $\alpha_\pm$ is on the left
hand side of the coordinate $x^i$.

The relation betweent the two representations can be also expresed in the
form:
\begin{equation}
b_\pm
=q^{3\over2}  [\alpha_+(\omega)-\alpha_-(\omega)]a_\pm{1\over
[\alpha_+(\omega)-\alpha_-(\omega)]}\ .
\label{ThD5}
\end{equation}

The operator $b^i_\pm$ is important when we investigate the transformation
rule of the creation-annihilation operator under the $*$-conjugation.

The Hamiltonian can be represented
in terms of the  creation-annihilation operator using the relations:

\begin{eqnarray}
(a_+\cdot a_-)
&=&q^{-{3\over2}}\lambda^{-1}\widehat{B}[-H(\omega)+Q_Nq^{N\over2}\alpha_-
(\omega)]\ ,\label{(3.15)}\\
(a_-\cdot a_+)
&=&q^{-{3\over2}}\lambda^{-1}\widehat{B}[-H(\omega)+Q_Nq^{N\over2}
\alpha_+(\omega)]\ .\label{(3.16)}
\end{eqnarray}
The linear combination of the above expressions gives the following
simple formula
\begin{equation}
(a_+\cdot a_-)+(a_-\cdot a_+)
=-q^{-{3\over2}}(1+q^N)\lambda^{-1}\widehat{B}H(\omega)\ , \label{(d.1)}
\end{equation}
where
\begin{equation}
\widehat{B}=1+{q^2-1\over\mu}(x\cdot\p)\ .
\label{(3.17)}
\end{equation}
The extra operator $\lambda^{-1}\widehat{B}$ commutes with the Hamiltonian:

\begin{equation}
[\lambda^{-1}\widehat{B},H(\omega)]=0\ .
\label{(3.18)}
\end{equation}
Actually the eigenfunctions also form a diagonal basis with respect to this
operator, i.e. when acting with the operator
$\lambda^{-1}\hat B$ onto the wave function eq.(\ref{(P1)}) we obtain
\begin{equation}
\lambda^{-1}\hat B\big|\Psi_p\big>\,={\cal N}_p\big|\Psi_p\big> \
,\label{(3.20)}
\end{equation}
where ${\cal N}_p$ is a certain $q$-number. However, this eigenvalue ${\cal
N}_p$ is not independent of the states.  By using the explicit form of the
wave function $\Psi_{p+2r}=S_pf_r(x^2)$ derived in section 2 we can
determine the eigenvalue ${\cal N}_p$.
It is given by
\begin{equation}
\lambda^{-1}\hat B\big|\Psi_{p+2r}\big>\,={q\over \mu q^{N\over
2}}(q^{-{N\over
2}-p+1}+q^{{N\over 2}+p-1})\big|\Psi_{p+2r}\big> \ .\label{EigenValueOfNp}
\end{equation}
{}From this we see that the eigenvalue of $\lambda^{-1}\hat B$ depends only on
the tensor structure defined by $p$ and is nonzero. To get the Hamiltonian,
 we have to divide out this operator.

\bigskip

We also can derive the expression of antisymmetric product of $a_+^i$ and
$a_-^j$:
\begin{equation}
{\cal P}_A(a_+^ia_-^j)
=\lambda^{-1}q^{N-3\over 2}{\cal
P}_A(x^i\p^j)[\alpha_+(\omega)-q^2\alpha_-(\omega)] \ . \label{(3.22)}
\end{equation}
The operator ${\cal P}_A(x^i\p^j)$ appearing in the r.h.s. is proportional
to the angular momentum in the limit $q\rightarrow1$.
The relations shown above suggest that the algebra of creation-annihilation
operators closes by including the angular momentum operator.
For this point see also ref.\cite{[Fio3]}.

Concerning the Hamiltonian it is not completely straightforward to express
it in
terms of the creation-annihilation operators as we see from the result of
eq.(\ref{(d.1)}).
One way to investigate such a property is to
 take the operators $a^i_{\pm}$ as the fundamental quantities
of the system, and consider the 'improved' Hamiltonian which is directly
proportional to $((a_+\cdot a_-)+(a_-\cdot a_+))$ on operator level. For
this one may still consider the rescaling of the creation-annihilation
operator by the function of the Hamiltonian as is suggested by
 the theorem D.  From eq.(\ref{ThD}), we see that when we define the
 creation-annihilation operator with the factor
 ${1\over\sqrt{K^2[H(\omega)]^2+4\omega^2}}$ appropriately, the relation
 between the improved $a^i_\pm$ and $b^i_\pm$ is simplified. Such an
 improved   creation-annihilation operator also seems to simplify the
 $*$-conjugation of the operators.

\section{Discussion}

We have presented the construction of the differential calculus on the
$q$-deformed Eucledian space.  There is no problem to introduce the reality
condition for the algebra generated only by the coordinate functions $x^i$.
However,
the problem of the reality condition for the algebra including
the differential operators is not completely settled.  There is a proposal
as we explained, for the algebra generated by $x^i$, $\p^i$ and
$\Lambda^{-1}$ \cite{[OZ]}.

We have given two different methods to construct the
solution of the $q$-deformed Schr\"odinger equation.  It is
explained how we can obtain the creation-annihilation operator which
generates all excitation levels. We also presented the $q$-analytic method
to derive the solutions in terms
of the $q$-polynomial and $q$-exponential functions by solving the
associated $q$-difference equation.

Still, we have to clarify the conjugation property, since under
the conjugation proposed in ref.\cite{[OZ]}, the Hamiltonian is not
an hermitian operator.
We considered the various possible Laplacians and its
operations on the $q$-exponential function, and found that
 $(\p\cdot\p)$, $\lambda^{-1}(\bar\p\cdot\p)$, $\lambda(\p\cdot\bar\p)$  and
$(\bar\p\cdot\bar\p)$ have simple relations which can be identified with the
eigenfunction equation.
These four
Laplacians are all non-hermitian under the $*$-conjugation
of ref.\cite{[OZ]}.

On the other hand, the remarkable point is that
the Hamiltonian discussed here has real energy eigenvalues. It means that
 we can solve the equation for $H^*(\omega)$ and we find the eigenfunctions
$\Psi^*$
with the same eigenvalue
 and tensor structure as for the case of $H(\omega)$.
There is even a unique correspondence between $\Psi$ and $\Psi^*$
\cite{[Fio2]}.
This properties may be a hint to reconsider the reality condition
for the $q$-deformed case, i.e., rather one should stay in the
complexified algebra with
trivial $*$-structure, and
impose a reality condition on the observables which
 can be specified by the $q$-correcepondence principle.
A proposal on this point is given in ref.\cite{[Fio2]}.
See also ref.\cite{[Wolfgang]}.

Finally, let us remark that
there is another interesting approach which is independent but possibly
related
to our approach. There, the quantum space algebra is identified with the
complex coordinate of the
Bargmann-Fock space \cite{[Achim],[Achim1],[Demichev]}.
\bigskip

\eject
\noindent {\Large\bf Acknowledgement}

The authors would like to acknowledge Prof. R. Sasaki for
inviting us to contribute the paper to this proceedings.
U.C. would like to acknowledge Prof. H. R\"omer for his hospitality
at Freiburg University, and
S.W. would like to thank Prof. K. Osterwalder for his hospitality at ETH.

\end{document}